\documentclass[dvips,10pt,prd,aps,twocolumn]{revtex4}
\usepackage{amsmath}
\usepackage{amsfonts}
\usepackage{graphicx}

\begin{document}

\title{General Relativity from a gauged WZW term.}
\author{Andr\'{e}s Anabal\'{o}n}
\email{anabalon-at-cecs.cl} \affiliation{ Centro de Estudios
Cient\'{\i}ficos (CECS) Casilla 1469 Valdivia, Chile}
\affiliation{Departmento de F\'{\i}sica, Universidad de
Concepci\'{o}n Casilla 160-C, Concepci\'{o}n, Chile}
\author{ Steven Willison}
\email{steve-at-cecs.cl}
\affiliation{\textbf{\textit{$^{1}$}}{\small Centro de Estudios
Cient\'{\i}ficos (CECS) Casilla 1469 Valdivia, Chile}}
\author{Jorge Zanelli}
\email{jz-at-cecs.cl} \affiliation{\textbf{\textit{$^{1}$}}{
Centro de Estudios Cient\'{\i}ficos (CECS) Casilla 1469 Valdivia,
Chile}}

\begin{abstract}
In this paper two things are done. First it is shown how a four dimensional gauged
Wess-Zumino-Witten term arises from the five dimensional Einstein-Hilbert plus Gauss-Bonnet
lagrangian with a special choice of the coefficients. Second, the way in which the equations of
motion of four-dimensional General Relativity arise is exhibited.
\end{abstract}
\maketitle

\section{Introduction}

Since the proof of power counting non renormalizability \cite{'tHooft:1974bx} of General
Relativity (\textbf{GR}), the scientific community has increasingly accepted the idea that
Einstein's theory is an effective field theory \cite{Burgess:2003jk}. Although some candidates for
the high energy limit of GR have emerged, no one has managed to give a power counting
renormalizable theory that also reproduces the dynamical behavior of GR in four dimensions.

The most general gravitational action in five dimensions has one additional free parameter besides
the cosmological constant \cite{Lovelock:1971yv}. These two parameters can be chosen so that the
lagrangian becomes a Chern-Simons form, acquiring some of the features that make $D=3$ gravity a
theory with zero beta function \cite{Witten:1988hc}. This particular $D=5$ gravitation theory is a
gauge theory where the spin connection $\omega $ and the vielbein $e$ are parts of a single
connection for the Lie algebras $so(4,2)$, $so(5,1)$ or $iso(4,1)$. Gravitation theories that have
these properties exist in all odd dimensions. They have been studied in
\cite{Chamseddine:1989nu,Mueller-Hoissen:1990vf} and a review of them can be found in Ref.
\cite{Zanelli:2005sa}. Another compelling reason to consider these gauge theories for the
$SO(D-1,2)$ group, they admit an immediate supersymmetric extension
\cite{Banados:1996hi,Troncoso:1998ng}. Moreover, the local supersymmetry in those theories is
realized off shell and without invoking auxiliary fields or ad-hoc constraints.

In this work, the topological sector --in the sense that no metric is needed to construct it--, of
the gauged Wess-Zumino-Witten (\textbf{gWZW}) lagrangian, is shown to arise from
higher-dimensional gravity. In the case where the connection is valued in the Lie algebra
$su_{L}(3)\times su_{R}(3)$, the gWZW term plus a kinetic piece for the Goldstone fields, arises
as the effective lagrangian of QCD \cite{Witten:1983tw}. In our case, the gauge group will not be
$SU_{L}(3)\times SU_{R}(3)$, but the anti-de-Sitter group in five dimensions, $SO(4,2)$.

\section{Five-dimensional gravity and gWZW terms}

The most general five dimensional, ghost-free \cite{Zumino:1985dp,Zwiebach:1985uq}, gravitational
action is given by\footnote{Throughout this work the exterior product between forms is not written
explicitly, i.e. $\omega \wedge e=\omega e$. Lower case Latin indices $a,b,c$ take values
$0...4.$, while capital indices $A,B,C$ cover the range $0...5.$.}
\begin{gather}
S(\omega ,e)=\int_{M}\varepsilon _{abcde}\big( \alpha _2 R^{ab}R^{cd}e^{e} +
\alpha_1 R^{ab}e^{c}e^{d}e^{e}  \notag \\
\hspace{2in}+ \alpha_0 e^{a}e^{b}e^{c}e^{d}e^{e}\big) ,
\end{gather}
where the curvature two-form is written in terms of the Lorentz (spin)connection $\omega$, as
\begin{equation}
R^{ab}=d\omega ^{ab}+\omega _{\ c}^{a}\omega ^{cb}=\frac{1}{2}R_{\text{ \ \ }%
\mu v}^{ab}dx^{\mu }dx^{v}
\end{equation}
The vielbein $e^{a}$ is related to the spacetime metric through $g_{\mu \nu}=e_{\mu }^a e_{v}^b
\eta_{ab}$, and $\eta_{ab}=diag(-,+,+,+,+)$ is the Lorentz-invariant metric. The vielbein and the
Lorentz connection are regarded as independent fields. The field equations associated with the
variations of $\omega $ are satisfied if the torsion, $De^{a}\equiv de^{a}+ \omega_{\text{ }b}^a
e^b$, is set equal to zero. In the sector of the theory where the torsion is zero and the vielbein
is invertible, $\omega $ is a function of the vielbein, and the usual second order equations for
the metric are recovered from the field equations obtained from the variation with respect to $e$.

An interesting accident occurs when the constants in the action are in the ratio $\alpha
_{2}:\alpha _{1}:\alpha_{0}=1:2/3:1/5$. In that case, the action can be rewritten as a
Chern-Simons theory
\cite{Chamseddine:1989nu,Mueller-Hoissen:1990vf,Banados:1996hi,Troncoso:1998ng},
\begin{align}
S(\mathcal{A})& =\kappa \int_{M}\left\langle \mathcal{A}d\mathcal{A}d%
\mathcal{A}+\frac{3}{2}\mathcal{A}^{3}d\mathcal{A}+\frac{3}{5}\mathcal{A}%
^{5}\right\rangle  \notag \\
& =\kappa \int_{M}CS(\mathcal{A)},  \label{CS}
\end{align}%
where
\begin{equation}
\mathcal{A}=\frac{1}{2}\omega ^{ab}J_{ab}+e^{a}J_{a5},\qquad \left\langle
J_{ab}J_{cd}J_{e5}\right\rangle =\varepsilon _{abcde},
\end{equation}%
$\kappa $ is dimensionless, $\left[ J_{AB},J_{CD}\right] =-J_{AC}\eta
_{BD}+J_{BC}\eta _{AD}-( A \leftrightarrow B)$, and $\langle ...\rangle $
stands for an invariant symmetric trace in the algebra.

In this way, the action acquires an enlarged gauge symmetry. If $\eta_{55}>0$ the gauge group is
$SO(5,1)$ and the action has positive cosmological constant. For $\eta_{55}<0$ the gauge group is
$SO(4,2)$ and the cosmological constant is negative. In the latter case, localized deformations of
the geometry give rise to asymptotically locally anti-de Sitter geometries.

Strictly speaking, under a gauge transformation the action (\ref{CS}) is not gauge invariant but
changes by a closed form plus a boundary term. This quasi invariance is a source of ambiguities in
an asymptotically $AdS$ spacetime, where the boundary terms that arise by gauge transformations
change the action and modify the conserved charges, producing even divergent values for them. This
problem can be circumvented if the action principle is modified by the addition of some new terms
that do not modify the field equations but render the action truly gauge invariant
\cite{Mora:2003wy,Mora:2006ka}. The trick is to replace the lagrangian in (\ref{CS}) by a
transgression form,
\begin{equation}
S(\mathcal{A},\bar{\mathcal{A}})=\kappa \int_{M}CS(\mathcal{A})-CS(\bar{%
\mathcal{A}})+\kappa \int_{\partial M}B(\mathcal{A},\bar{\mathcal{A}}),
\label{TPP}
\end{equation}%
where
\begin{gather}
B(\mathcal{A},\bar{\mathcal{A}})= -\Big\langle \mathcal{A} \bar{\mathcal{A}}
\left(\mathcal{F} + \bar{\mathcal{F}} - \frac{1}{2}\mathcal{A}^2 -\frac{1}{2}%
\bar{\mathcal{A}}^2+\frac{1}{2}\mathcal{A}\bar{\mathcal{A}} \right)%
\Big\rangle.
\end{gather}

The transgression form is the object which appears in the Chern-Weil theorem, that states that the
pullback of invariant polynomials of the curvature $P(\mathcal{F})$ are members of cohomology
groups of the manifold where they are defined \cite{Nakahara:1990th},
\begin{equation}
dP(\mathcal{F})=0,\qquad P(\mathcal{F})-P(\mathcal{\bar{F}})=dTP(\mathcal{A},%
\bar{\mathcal{A}}),  \label{TP}
\end{equation}%
where $TP(\mathcal{A},\bar{\mathcal{A}})$ is defined by equation (\ref{TP})
up to a closed form. The gauge invariant, globally-defined expression for $%
TP(\mathcal{A},\bar{\mathcal{A}})$ stands for the transgression form. In $%
2n-1$ dimensions, the transgression takes the form
\begin{equation}
TP_{2n-1}(\mathcal{A},\bar{\mathcal{A}})=n\int_{0}^{1}dt\ \left\langle\left(
\mathcal{A}-\bar{\mathcal{A}}\right) \mathcal{F}_{t}^{n-1}\right\rangle,
\end{equation}
where $\mathcal{F}_{t}=d\mathcal{A}_{t}+\mathcal{A}_{t}\mathcal{A}_{t}$, $%
\mathcal{A}_{t} = \mathcal{A}\left( 1-t\right) + \bar{\mathcal{A}}t$. Thus, the boundary term
$B(\mathcal{A},\bar{\mathcal{A}})$ is uniquely determined by the Chern-Weil theorem.

The field equations for $\mathcal{A}$ are the same, whether the action principle is defined by
(\ref{TPP}) or by (\ref{CS}). However, there is a problem interpreting the physical meaning of the
field $\bar{\mathcal{A}}.$ An interpretation was proposed in \cite{Mora:2006ka}, where the
lagrangian $CS(\bar{\mathcal{A}})$ was considered as defined in a manifold with opposite
orientation to the one for $CS(\mathcal{A})$. An alternative is to regard $\bar{\mathcal{A}}$ not
as a dynamical field but as a means of constructing the boundary term which makes the action
finite \cite{Miskovic+Olea}. A different philosophy will be followed here, that is to regard
$\bar{\mathcal{A}}$ and $\mathcal{A}$ as two connections defining the same non trivial, principal
bundle. That is, they are related by a gauge transformation.

If a non trivial bundle is considered, the integrand does not exist globally either in (\ref{CS})
or in (\ref{TPP}). This non existence problem will be treated in \cite{AWZ1}, where a detailed
study of the definition of an action principle in a manifold divided into patches, for a non
trivial principal bundle, will be presented \footnote{The $D=3$ case was studied in
\cite{Dijkgraaf:1989pz}}. In the case of a non trivial bundle, however, the action (\ref{TPP}) can
be treated formally provided more than one chart is used. In this case, it is necessary to
introduce connection one-forms defined on each chart, such that, in the overlap of two charts the
connections are related by a gauge transformation,
$\bar{\mathcal{A}}=h^{-1}\mathcal{A}h+h^{-1}dh\equiv \mathcal{A}^{h}$, where $h$ is a transition
function which determines the non triviality of the bundle.

Replacing $\bar{\mathcal{A}}= \mathcal{A}^{h}$ in (\ref{TPP}), it is straightforward to check that
the action takes the form of a gWZW term,
\begin{widetext}
\begin{eqnarray}
&&S(h,\mathcal{A})=-\frac{\kappa}{10}\int_{M^5}\left\langle h^{-1}dh
  (h^{-1}dh)^2
 (h^{-1}dh)^2\right\rangle
 +\kappa\int_{M^4} \left \langle dhh^{-1} {\cal A} \left( d\mathcal{ A}
 +
 \frac{1}{2}{\cal A}^2 \right)\right\rangle
 \label{WZ} \\&&\notag
 -\frac{\kappa}{2} \int_{M^4} \left\langle dh h^{-1} \mathcal{A}
 \left\{(dhh^{-1})^2 + {\cal A}\, dhh^{-1} \right\}\right\rangle
 - \kappa\int_{M^4}\left\langle {\cal A}  {\cal A}^h
 \left( \mathcal{ F} + \mathcal{ F}^h -\frac{1}{2}
 {\cal A}^2  -\frac{1}{2}({\cal A}^h )^2 + \frac{1}{2}{\cal A} {\cal A}^h\right) \right\rangle\,
 .
\end{eqnarray}
\end{widetext}
where the curvature is $\mathcal{F}=d\mathcal{A}+\mathcal{AA}$ and $\mathcal{%
F}^{h}= h^{-1}\mathcal{F}h.$ This action is invariant under the adjoint
action of the gauge group, namely,
\begin{equation}
\mathcal{A}\rightarrow g^{-1}\mathcal{A}g+g^{-1}dg,\qquad h\rightarrow
g^{-1}hg.
\end{equation}

As has been shown, the principle of gauge invariance, through the mathematical structure of the
theory of principal bundles, provides a compactification mechanism. Beginning with a five
dimensional gauge theory that has no metric in it, a $D=4$ gauge invariant theory has been
obtained. This is a compactification mechanism alternative to Kaluza-Klein. The relation between
three-dimensional Chern-Simons theories and two-dimensional gWZW models was early realized in
\cite{Moore:1989yh}. Originally, gWZW terms were obtained in \cite{Witten:1983tw} by trial and
error, and it was later shown that they can be obtained systematically see, e. g.,
\cite{Alvarez-Gaume:1984dr,Witten:1991mm,deAzcarraga:1998bu}.

Other attempts to relate the $D=4$ gWZW lagrangians to $D=5$ Chern-Simons theory can be found in
the literature (see for instance, refs. \cite{Banados-96,Gegenberg}). However, asymptotic
conditions for the metric were always assumed in order to reproduce the kinetic term for the
Goldstone fields, not present in (\ref{WZ}). As can been seen from the previous discussion, such a
strong assumption is not required here, the kinetic term does not appear, but the gWZW term arises
naturally from the transgression form.

The action (\ref{WZ}), usually supplemented with a kinetic term for the so-called Goldstone
fields, requires the introduction of the Hodge dual, which in turn requires the existence of a
metric in the manifold. The point of view followed here is that the metric arises from the
components of a gauge connection in a broken phase of the theory, but it is not assumed to be
defined a priori. In the next section the action principle (\ref{WZ}) for a particular class of
$h$ will be studied further.

\section{The gWZW term as a gravitational action}

The action (\ref{WZ}) describes a theory with $SO(4,2)\times SO(4,2)$ gauge symmetry spontaneously
broken to its diagonal subgroup $SO(4,2)$\footnote{For the sake of simplicity the discussion is
restricted to $G=SO(4,2),$ the extension of the results to $SO(5,1)$ is trivial. In this section
the indices $a,b$ take values in the range, $0,1,2,3$.}  \cite{Witten:1983tw}. The gauge
invariance is manifest since the connection takes its values in the diagonal subalgebra only
\cite{Witten:1991mm}. However, in order to describe the known low energy gravitational behavior,
it is necessary to reduce the symmetry to the usual $SO(3,1)$ local symmetry present in the
Einstein-Hilbert action. A way to do this is by replacing the gauge group $SO(4,2)$ by the coset
$\frac{SO(4,2)}{\mathbb{R}}$, where $R$ stands for the group of transformations generated by
$e^{\phi J_{45}}$ and an element of the coset, $h$, is a representative of the equivalence class
$\left[ h\right] =\left\{ h\sim h^{\prime }\Longleftrightarrow h^{\prime }=e^{\phi
J_{45}}h|h,h^{\prime }\in SO(4,2)\right\}$. Using the adjoint action, the stability group of this
coset is $SO(3,1)\times R$ and it corresponds to the residual gauge invariance present in the
theory, that is, $h\in \left[ h\right] \Longrightarrow g^{-1}hg\in \left[ h\right]
\Longleftrightarrow g\in SO(3,1)\times R$.

\begin{figure}[h]
  \includegraphics[width=.4\textwidth]{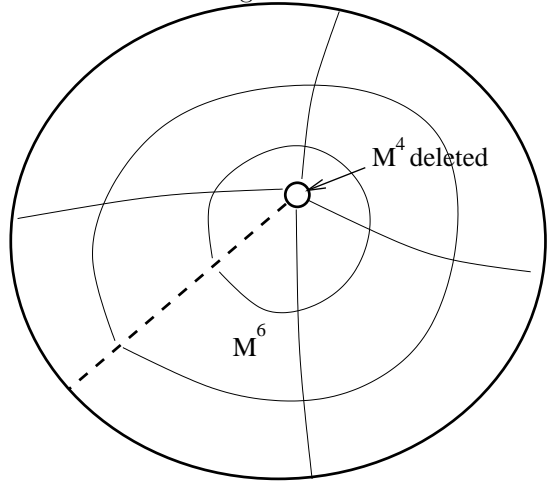}\\
  \caption{A four-dimensional defect in a six-dimensional manifold.
  The submanifold $M^4$ has been deleted, as indicated by the infinitesimal
  loop in the center of the diagram.}\label{6d_fig}
\end{figure}

Fixing the Goldstone field associated to $J_{45}$ in the action
(\ref{WZ}), corresponds to reducing the gauge symmetry down to
$SO(3,1) \times R$.  There is an interesting geometrical
interpretation of this. Suppose we have a six-dimensional manifold
$M^{6}$ and delete a four dimensional submanifold $M^{4}$ (Fig.
\ref{6d_fig}).

Then, by considering the integral of a characteristic class on $M^{6}-M^{4}$, the action induced
on $M^{4}$ is (\ref{WZ}) \cite{AWZ2},
\begin{equation}
\int_{M^{6}-M^{4}}\left\langle \mathcal{FFF}\right\rangle =S(h,\mathcal{A)}.
\end{equation}
The field $\phi$ can be interpreted in terms of the
six-dimensional pseudo-Riemannian geometry, as a deficit angle
around the four-dimensional defect $M^4$ which, as shown here, is
related to the four dimensional cosmological constant. Here we
have assumed that $\phi$ is a constant, thus breaking part of the
gauge symmetry ``by hand". In terms of this geometrical picture,
the defect is assumed to have a fixed deficit angle.

Now, in order to write the field equations associated with the
gWZW term (\ref{WZ}) for the coset $\frac{SO(4,2)}{\mathbb{R}}$,
it is helpful to decompose the connection and the curvature in a
way that reflects the $SO(3,1)$ symmetry,
\begin{eqnarray}
\mathcal{A} &=&\frac{1}{2}\omega ^{ab}J_{ab}+b^{a}J_{a4}+e^{a}J_{a5}+\Phi J_{45}, \\
\nonumber \mathcal{F} &=&\frac{1}{2}(R^{ab}+e^{a}e^{b}-b^{a}b^{b})J_{ab}+[Db^{a}+ e^a \Phi]J_{a4}
\\
&&+[De^{a}+b^a\Phi]J_{a5}+[d\Phi-b_{a}e^{a}]J_{45},
\end{eqnarray}%
where $De^{a}=de^{a}+\omega_{\text{ }b}^{a}e^{b}$. The field equations associated to the variation
of the Goldstone fields in the coset $\frac{SO(4,2)}{\mathbb{R}}$ are
\begin{widetext}\begin{gather} \kappa
\int_{M^{4}}\Big\langle h^{-1}\delta h\Big\{{}\!\left(
\mathcal{F}^{h}\right) ^{2}+\mathcal{F}^{2}+{}\!\mathcal{F}^{h}\mathcal{F}-
\frac{3}{4}[{}\!\mathcal{A}^{h}-\mathcal{A},{}\mathcal{A}^{h}-\mathcal{A}]\
({}\!\mathcal{F}^{h}+\mathcal{F})  \notag \\ \hspace{2in}
+\frac{1}{8}[{}\!\mathcal{A}^{h}-\mathcal{A},{}\!\mathcal{A}^{h}-\mathcal{A}]^{2}+
\frac{1}{2}({}\!\mathcal{A}^{h}-\mathcal{A})[{}\!\mathcal{F}^{h}+
\mathcal{F},{}\!\mathcal{A}^{h}-\mathcal{A}])\Big\}\Big\rangle=0,  \label{2}
\end{gather}\newline
while the $15$ equations of motion that arise from the variation of the connection, are
\begin{equation}
\kappa \int_{M^{4}}\langle \delta \mathcal{A} ({}\!\mathcal{A}^{h}-\mathcal{A} )\left(
{}\!\mathcal{F}^{h}+2\mathcal{F}-\frac{1}{4}[{}\!\mathcal{A}^{h}-
\mathcal{A},{}\!\mathcal{A}^{h}-\mathcal{A}]\right) \rangle -(h\leftrightarrow h^{-1})=0.
\label{con}
\end{equation}\end{widetext}

In order to obtain the equations of motion in a explicit form, it is necessary to pick a
representative of the coset. In any open set, it can be parametrized by $14$ coordinates $\pi $,
as $h=e^{\phi J_{45}}e^{\pi ^{a4}J_{a4}}e^{\pi ^{a5}J_{a5}}e^{\frac{1}{2}\pi ^{ab}J_{ab}}$, where
$\phi $ is an arbitrary, real, constant. The purely gravitational sector of the theory, that is
the one in which only the vielbein and the spin connection are present, corresponds to setting
$b^{a}$, $\Phi $ and the $14$ Goldstone fields equal to zero. In this dynamical sector the set of
equations (\ref{con}) reduces to
\begin{eqnarray}
\varepsilon _{abcd}\;e^{b}\left( R^{cd}+\mu e^{c}e^{d}\right)
\sinh \phi &=&0  \label{EH} \\
\varepsilon _{abcd}\;e^{c}De^{d}\sinh \phi &=&0.  \label{TOR}
\end{eqnarray}

Here the constant $\mu$ is given by $\left( 1+2\cosh \phi \right)/3$. Excluding the trivial case
$\phi=0$ implies $\mu >1$, and equation (\ref{TOR}) implies that the torsion is zero. Solving the
torsion for the spin connection and replacing it back in (\ref{EH}), the Einstein's equations in
standard form are obtained. It is reassuring to check that these field configurations also satisfy
the $14$ field equations obtained from (\ref{2}).

In order to write Einstein's equations it is necessary to assume that the vielbein $e_{\text{ }\mu
}^a$ is invertible, and to make contact with a metric theory, it is also necessary to rescale the
vielbein with a parameter with dimensions of length, $e^{a}=\bar{e}^{a}/l$. In this way, the
metric is $g_{\mu v}=\bar{e}_{\mu }^{a}\bar{e}_{v}^{b}\eta _{ab}$, the effective cosmological
constant acquires its usual units $\Lambda =\left( 1+2\cosh \phi \right)l^{-2}$, and (\ref{EH})
can be rewritten as
\begin{equation}
R_{\mu v}-\frac{1}{2}g_{\mu v}R-\Lambda g_{\mu v}=0.
\end{equation}

However, as was shown in ref. \cite{Witten:1988hc}, assuming the invertibility of the vielbein
spoils the possibility of making a sensible, quantum mechanical, perturbative expansion around
$\mathcal{A}=0$. The rescaling of $e^a$ is also unhelpful in the sense that, if no such rescaling
is done, all the parameters of the theory are dimensionless, which would suggest the possibility
of power counting renormalizability of the theory.

\section{Discussion and Outlook}

Here we have shown that General Relativity is a dynamical sector of a gWZW theory for the coset
$\frac{SO(4,2)}{\mathbb{R}}$. It can be checked that the same phenomenon occurs if in the
$SO(4,2)$ gWZW, a representative of the group is taken as $h=e^{\phi J_{45}}e^{\pi
^{a4}J_{a4}}e^{\pi ^{a5}J_{a5}}e^{\pi ^{ab}J_{ab}}=e^{\phi J_{45}}\bar{h}$, and $\phi$ is kept
fixed in the action reducing the symmetry to $SO(3,1)\times R$. The four-dimensional spacetime
arises in the sector of solution space characterized by $\bar{h}=1$, det$e\neq 0$, $b=0=
\mathcal{A}^{45}$.

Having obtained the equations of General Relativity, a more detailed analysis of the dynamical
structure of the theory (\ref{WZ}) is necessary. Generically, as it happens with all higher
dimensional Chern-Simons theories, the system will possess degenerate dynamical sectors
\cite{Saavedra:2000wk,Miskovic:2003ex,Miskovic:2005di}. A deeper understanding of this problem
would be required prior to any study of the quantum properties of the action (\ref{WZ}).

The theory changes dramatically if the $\phi$ field is regarded as dynamical. In that case, there
is no purely gravitational sector with only $e^a$ and $\omega^{ab}$ nonzero. However there are
solutions of gravity coupled to the other fields, including an interesting class of gravitational
solitons, that is, Lorentzian, everywhere regular, classical solutions. A two-parameter family of
these solitons will be presented in \cite{AWZ3}.

\textbf{Acknowledgements}

The authors wish to thank Eloy Ayon-Beato, Sara Farese, Gast\'{o}n
Giribet, Joaquim Gomis, Elias Gravanis, Julio Oliva, Tom\'{a}s
Ort\'{\i}n and Ricardo Troncoso for enlightning discussions. This
work has been supported in part by FONDECYT grants $N^{o}$s 1061291,
1060831, 1040921 and 3060016. A.A. wishes to thanks the support of
MECESUP UCO 0209 and CONICYT grants during the realization of this
work. Institutional support to the Centro de Estudios
Cient\'{\i}ficos (CECS) from Empresas CMPC is gratefully
acknowledged. CECS is funded in part by grants from the Millennium
Science Initiative, Fundaci\'{o}n Andes, the Tinker Foundation.

\end{document}